\documentclass{aastex61}

\date{\today}

\begin{document}

\title{Planck 2015 constraints on the non-flat XCDM inflation model}

\author{Junpei Ooba}
\altaffiliation{ooba.jiyunpei@f.mbox.nagoya-u.ac.jp}
\affiliation{Department of Physics and Astrophysics, Nagoya University, Nagoya 464-8602, Japan}

\author{Bharat Ratra}
\affiliation{Department of Physics, Kansas State University, 116 Cardwell Hall, Manhattan, KS 66506, USA}

\author{Naoshi Sugiyama}
\affiliation{Department of Physics and Astrophysics, Nagoya University, Nagoya 464-8602, Japan}
\affiliation{Kobayashi-Maskawa Institute for the Origin of Particles and the Universe, Nagoya University, Nagoya, 464-8602, Japan}
\affiliation{Kavli Institute for the Physics and Mathematics of the Universe (Kavli IPMU), The University of Tokyo, Chiba 277-8582, Japan}



\begin{abstract}

We examine the Planck 2015 cosmic microwave background (CMB) anisotropy data 
by using a physically-consistent energy density inhomogeneity power spectrum 
generated by quantum-mechanical fluctuations during an early epoch of 
inflation in the non-flat XCDM model. Here dark energy is parameterized using 
a fluid with a negative equation of state parameter but with the speed of
fluid acoustic inhomogeneities set to the speed of light. We find that the 
Planck 2015 data in conjunction with baryon acoustic oscillation distance 
measurements are reasonably well fit by a closed XCDM model in which spatial 
curvature contributes a percent of the current cosmological energy density 
budget. In this model, the measured non-relativistic matter density parameter 
and Hubble constant are in good agreement with values determined using most 
other data. Depending on cosmological parameter values, 
the closed XCDM model has reduced power, relative to the tilted, 
spatially-flat $\Lambda$CDM case, and can partially alleviate the low 
multipole CMB temperature anisotropy deficit and can help partially reconcile 
the CMB anisotropy and weak lensing $\sigma_8$ constraints, at the expense of 
somewhat worsening the fit to higher multipole CMB temperature anisotropy 
data. However, the closed XCDM inflation model does not seem to improve the 
agreement much, if at all, compared to the closed $\Lambda$CDM inflation case,
even though it has one more free parameter.
\end{abstract}

\keywords{cosmic background radiation --- cosmological parameters --- large-scale structure of universe  --- inflation --- observations}



\section{Introduction} \label{sec:intro}

We recently found that cosmic microwave background (CMB) anisotropy 
measurements do not require flat spatial hypersurfaces in the $\Lambda$CDM
scenario \citep{Oobaetal2018a, ParkRatra2018a, ParkRatra2018b}, provided one 
uses a physically consistent
non-flat model power spectrum of energy density inhomogeneities 
\citep{RatraPeebles1995, Ratra2017} in the analysis of the CMB data.

In the standard $\Lambda$CDM model \citep{Peebles1984}, dark energy,
taken to be the cosmological constant $\Lambda$, dominates the current 
cosmological energy budget and powers the currently accelerating 
cosmological expansion. Cold dark matter (CDM) is the next largest 
contributor to the current energy budget, followed by baryonic matter 
and small contributions from neutrinos and photons. The standard
$\Lambda$CDM model assumes flat spatial geometry. For reviews of the 
standard and related scenarios, see  \citet{RatraVogeley2008}, 
\citet{Martin2012}, \citet{Joyceetal2016}, \citet{HutererShafer2017} 
and references therein.

It is conventional to parameterize the standard flat-$\Lambda$CDM model in 
terms of six variables: $\Omega_{\rm b} h^2$ and $\Omega_{\rm c} h^2$, the 
current values of the baryonic and cold dark matter density parameters
multiplied by the square of the Hubble constant (in units of 100 km s$^{-1}$ 
Mpc$^{-1}$); $\theta$, the angular diameter distance as a multiple of the 
sound horizon at recombination; $\tau$, the reionization optical depth; 
and $A_{\rm s}$ and $n_{\rm s}$, the amplitude and spectral index of the
(assumed) power-law primordial scalar energy density inhomogeneity power 
spectrum, \citep{Adeetal2016a}. The predictions of the flat-$\Lambda$CDM 
model are largely consistent with most available observational constraints 
\citep[][and references therein]{Adeetal2016a}.    

There are suggestions that flat-$\Lambda$CDM might not be 
as compatible with more recent, larger compilations of measurements  
\citep{Solaetal2017a, Solaetal2017b, Solaetal2018, Solaetal2017c, Zhangetal2017}
that might be more consistent with dynamical dark energy models. These 
include the simplest, physically consistent, seven parameter 
flat-$\phi$CDM model in which a scalar field $\phi$ with potential energy 
density $V(\phi) \propto \phi^{-\alpha}$ is the dynamical dark energy 
\citep{PeeblesRatra1988, RatraPeebles1988} and $\alpha > 0$ is the seventh 
parameter and controls dark energy evolution. 

Compared to the time-independent $\Lambda$, dynamical dark energy density 
evolves in manner closer to (but slower than) that of spatial curvature 
energy density. This can cause a complication. Under the assumption of
spatial flatness, CMB anisotropy measurements favor $\Lambda$ although mild 
dark energy time evolution remains viable. When the general, non-flat, 
$\Lambda$CDM model is used to analyze the CMB data they indicate that 
spatial hypersurfaces are close to flat although mildly curved space 
is acceptable. 
When CMB anisotropy observations are analyzed in a dynamical dark energy model
with non-flat spatial hypersurfaces there is degeneracy between spatial 
curvature and the parameter governing dark energy dynamics. This results in 
weaker constraints on both parameters, when compared to the cases where 
only non-zero spatial curvature or only dynamical dark energy is assumed
\citep[early references include][and references therein]{AurichSteiner2002, AurichSteiner2003, AurichSteiner2004, Crooksetal2003, IchikawaTakahashi2006, Wright2006, Ichikawaetal2006, Zhaoetal2007, IchikawaTakahashi2007, Clarksonetal2007, WangMukherjee2007, Gongetal2008, IchikawaTakahashi2008, Hlozeketal2008, Vireyetal2008}  

Non-zero spatial curvature brings in a new length scale, in addition to 
the Hubble scale. Consequently in the non-flat
case one must use an inflation model to compute a consistent power spectrum.
For open spatial hypersurfaces the \citet{Gott1982} open-bubble inflation 
model is taken as the initial epoch of the cosmological model and one 
computes zero-point quantum fluctuations during the open inflation epoch and 
propagates these to the current open, accelerating cosmological expansion 
epoch where they are 
energy density inhomogeneities \citep{RatraPeebles1994, RatraPeebles1995}.\footnote{Observational consequences of the open inflation model are discussed in 
\citet{Kamionkowskietal1994}, \citet{Gorskietal1995}, \citet{Gorskietal1998}, \citet{Ratraetal1999}, and 
references therein.}
For closed spatial hypersurfaces Hawking's prescription for the quantum state 
of the universe \citep{Hawking1984} can be used to construct a closed 
inflation model \citep{Ratra1985, Ratra2017}.
The initial closed inflation epoch linear perturbation solutions constants 
of integration are determined from closed de Sitter 
invariant quantum mechanical initial conditions in the Lorentzian section 
of the closed de Sitter space that follow from Hawking's prescription 
that the quantum state of the universe only include field configurations 
regular on the Euclidean (de Sitter) sphere sections \citep{Ratra1985, Ratra2017}. These initial conditions are the unique de Sitter invariant ones. 
Zero-point quantum-mechanical fluctuations during closed inflation 
provide a late-time energy density inhomogeneity power spectrum that is not 
a power law \citep{Ratra2017}; it is a generalization to the 
closed case \citep{WhiteScott1996, Starobinsky1996, Zaldarriagaetal1998, Lewisetal2000, LesgourguesTram2014} of the flat-space scale-invariant 
spectrum.

We recently analyzed the Planck 2015 CMB anisotropy data in the non-flat
$\Lambda$CDM inflation model, using the consistent power spectrum for the 
non-flat case \citep{Oobaetal2018a}.\footnote{A similar analysis, with similar conclusions, has been carried out for the non-flat $\phi$CDM inflation model \citep{Oobaetal2018b, ParkRatra2018c}.}
In this paper we examine what constraints 
these data place on a model in which dark energy is parameterized in terms of 
a fluid,
the so-called XCDM parameterization. Here the dark energy fluid pressure and 
energy density are related via $p_X = w_0 \rho_X$, where $w_0$ is the equation 
of state parameter and is taken to be $< -1/3$. It is well known that such 
a fluid is unstable and to stabilize it we arbitrarily require that spatial 
inhomogeneities in the fluid propagate at the speed of light.  

Compared to the six parameter flat-$\Lambda$CDM inflation model discussed 
above, in the non-flat case there is no simple tilt option, so 
$n_{\rm s}$ is no longer a free parameter and is replaced by the current value 
of the curvature density parameter $\Omega_{\rm k}$ which results in 
the six parameter non-flat $\Lambda$CDM model \citep{Oobaetal2018a}. Here
we replace the cosmological constant $\Lambda$ by the dark energy fluid 
parameterized by $w_0$, which is a new free parameter, resulting in the 
seven parameter non-flat XCDM inflation model.\footnote{We study the seven 
parameter flat XCDM inflation model elsewhere \citep{Oobaetal2018c}.}
Of course, this is not 
a physical model, and it is also unable to properly mimic the dark energy 
evolution of the scalar field dynamical dark energy $\phi$CDM model 
\citep{PodariuRatra2001}. 
Nevertheless, it is a simple parameterization of dynamical dark energy
that is relatively straightforward to use in a computation and that is 
worth exploring as a first, and hopefully not misleading, attempt to gain 
some insight into how dark energy
dynamics and non-zero spatial curvature jointly influence the CMB anisotropy 
data constraints.  We note here that there is a physically consistent 
non-flat seven parameter scalar field dynamical dark energy 
model \citep{Pavlovetal2013}; because the scalar field inhomogeneities must
be accounted for in this model, the computational demands in $\phi$CDM are 
much more significant than in the XCDM case we study here.   

In this paper we use the Planck 2015 CMB anisotropy data to constrain this 
seven parameter non-flat XCDM inflation model.\footnote{\citet{Zhaoetal2017} and \citet{DiValentinoetal2017} have recently studied other variants of flat XCDM designed to resolve cosmological tensions.}
We find in this model that 
the Planck 2015 CMB anisotropy data used jointly with baryon acoustic 
oscillation (BAO) distance measurements do not require that spatial 
hypersurfaces be flat. The data favor a slightly closed model. These results 
are consistent 
with our earlier analysis of the six parameter non-flat $\Lambda$CDM 
inflation model \citep{Oobaetal2018a}.

In our analyses here we use different CMB anisotropy data combinations
\citep{Adeetal2016a}. For CMB data alone, we find that the best-fit non-flat
XCDM model, for the TT + lowP + lensing Planck 2015 data, has spatial 
curvature density parameter $\Omega_{\rm k} = - 0.021 {+ 0.023 \atop - 0.005} {+0.029 \atop - 0.050}$ (1 and 2$\sigma$ error bars) and is mildly closed. When 
we include the BAO data that Planck 2015 used, we find for the TT + lowP + 
lensing CMB anisotropy case that $\Omega_{\rm k} = - 0.008 \pm 0.003 \pm 0.006$.

As with the six parameter closed-$\Lambda$CDM models, it might be significant 
that the best-fit seven parameter closed XCDM models have less low $\ell$ 
CMB temperature anisotropy $C_\ell$ power than does the best-fit six parameter 
tilted, spatially-flat $\Lambda$CDM model, and so appear to be
in slightly better agreement with the low-$\ell$ temperature $C_\ell$ 
observations (less so when the BAO data are included in the analysis). Overall, 
however, the best-fit seven parameter closed XCDM inflation model does not 
do better, and probably does a little worse, than the best-fit six parameter 
closed $\Lambda$CDM inflation model we studied earlier \citep{Oobaetal2018a}.  

In Sec.\ II we summarize the methods we use. Our parameter constraints are 
plotted, tabulated, and discussed in Sec.\ III, where we also attempt to 
determine how well the best-fit closed-XCDM case fits the data. Conclusions
are given in Sec.\ IV.

\section{Methods}

\begin{figure}[ht]
\plotone{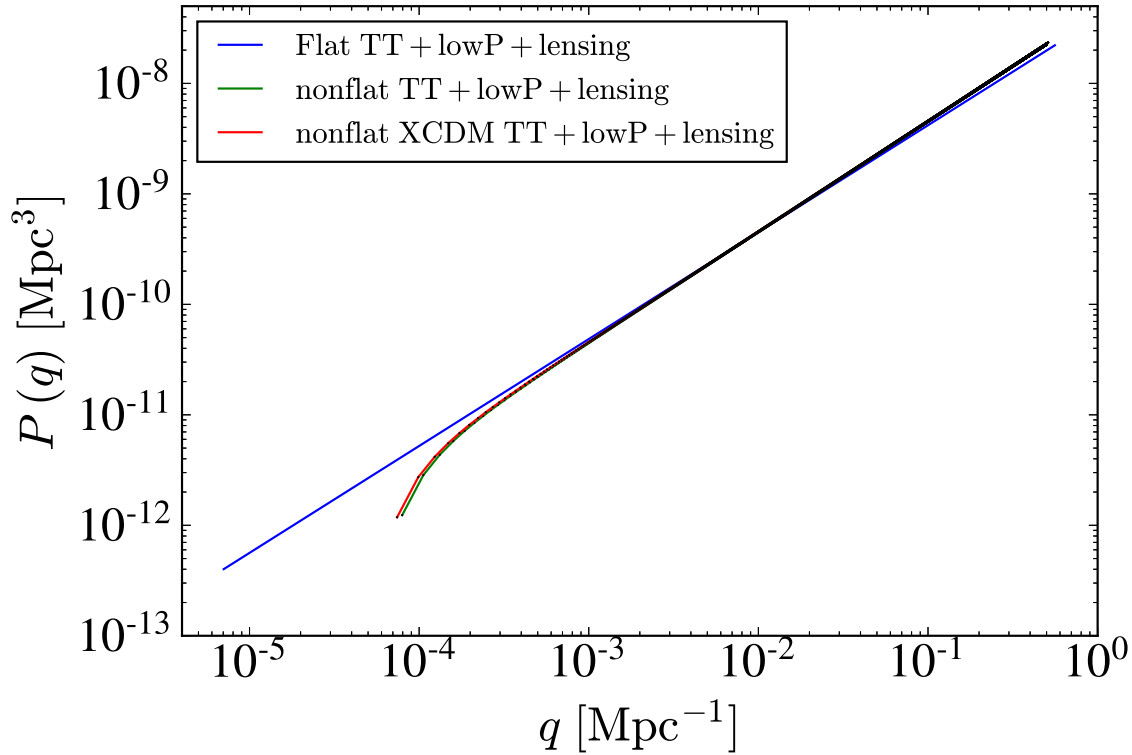}
\caption{Best-fit (see text) gauge-invariant fractional energy density 
inhomogeneity power spectra. The blue line corresponds to the tilted
flat-$\Lambda$CDM model of \citet{Adeetal2016a}. In the closed case, 
wavenumber $q \propto A + 1$ where the eigenvalue 
of the spatial Laplacian is $-A(A+2)$, $A$ is a non-negative integer, 
$A = 0$ corresponds to the constant zero-mode on the three sphere, the power 
spectrum vanishes at $A = 1$, and the points on the red and green curves 
correspond to $A = 2, 3, 4, ...$, see eqns.\ (8) and (203) of 
\citet{Ratra2017}. On large scales the power spectra for the best-fit 
closed XCDM (red curve) and $\Lambda$CDM (green curve) models are
suppressed relative to that of the best-fit tilted flat-$\Lambda$CDM model.
The $P(q)$ are normalized using the best-fit values of $A_{\rm s}$ at the 
pivot scale $k_0 = 0.05$ [Mpc${}^{-1}$]. 
\label{fig:Pk}}
\end{figure}

For our non-flat model analyses we use the open and closed 
inflation model energy density inhomogeneity power spectrum
\citep{RatraPeebles1995, Ratra2017}. Figure \ref{fig:Pk} compares 
closed XCDM and $\Lambda$CDM inflation model power spectra and a tilted 
flat-$\Lambda$CDM inflation model power spectrum. We use the public numerical code 
CLASS \citep{Blasetal2011} to compute the angular power spectra of the CMB 
temperature, polarization, and lensing potential anisotropies. Our parameter 
estimations are carried out using the Markov chain Monte Carlo method program 
Monte Python \citep{Audrenetal2013}.

We assume flat priors for the cosmological parameters over the ranges
\begin{eqnarray}
\label{eq:prior}
&100\theta \in (0.5,10),\ \ \Omega_{\rm b}h^2 \in (0.005,0.04),\ \ \Omega_{\rm c}h^2 \in (0.01,0.5), \nonumber\\
&\tau \in (0.005,0.5),\ \ {\rm ln}(10^{10}A_{\rm s}) \in (0.5,10),\ \ \Omega_{\rm k} \in (-0.5, 0.5),\ \ w_0 \in (-3, 0.2).
\end{eqnarray}
The CMB temperature and the effective number of neutrinos are taken to be 
$T_{\rm CMB}= 2.7255\ \rm K$ from COBE \citep{Fixsen2009} and $N_{\rm eff}=3.046$ 
with one massive (0.06 eV) and two massless neutrino species in a normal 
hierarchy. The primordial helium fraction $Y_{\rm He}$ is 
inferred from standard Big Bang nucleosynthesis as a function of the baryon 
density.

We constrain model parameters by comparing model predictions to the CMB angular 
power spectrum data from the Planck 2015 mission \citep{Adeetal2016a} and the 
BAO distance measurements of the 6dF Galaxy Survey \citep{Beutleretal2011}, 
the Baryon Oscillation Spectroscopic Survey (LOWZ and CMASS) 
\citep{Andersonetal2014}, and the Sloan Digital Sky Survey main galaxy sample 
(MGS) \citep{Rossetal2015}.

\section{Results}

\begin{figure}[ht]
\plottwo{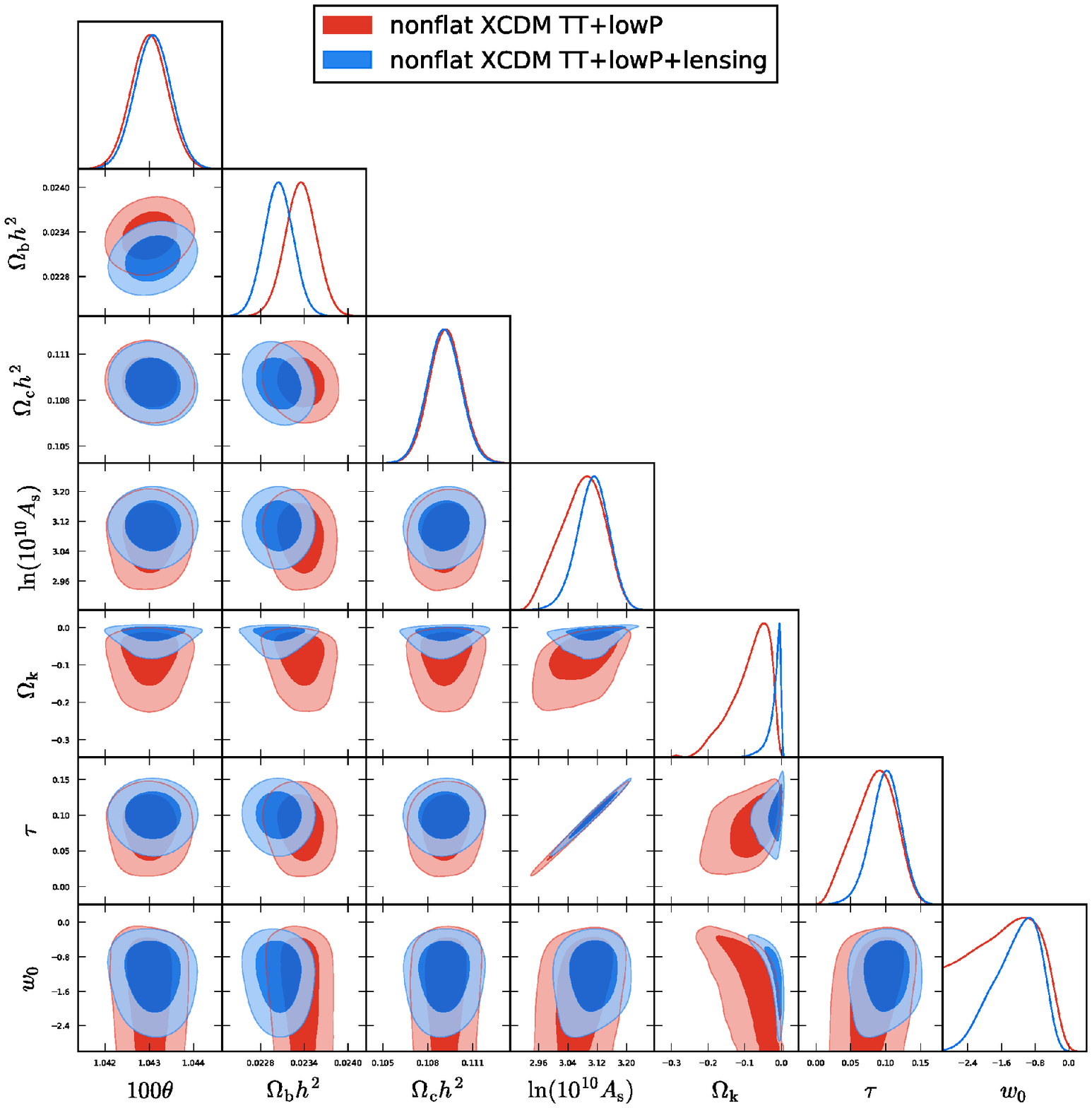}{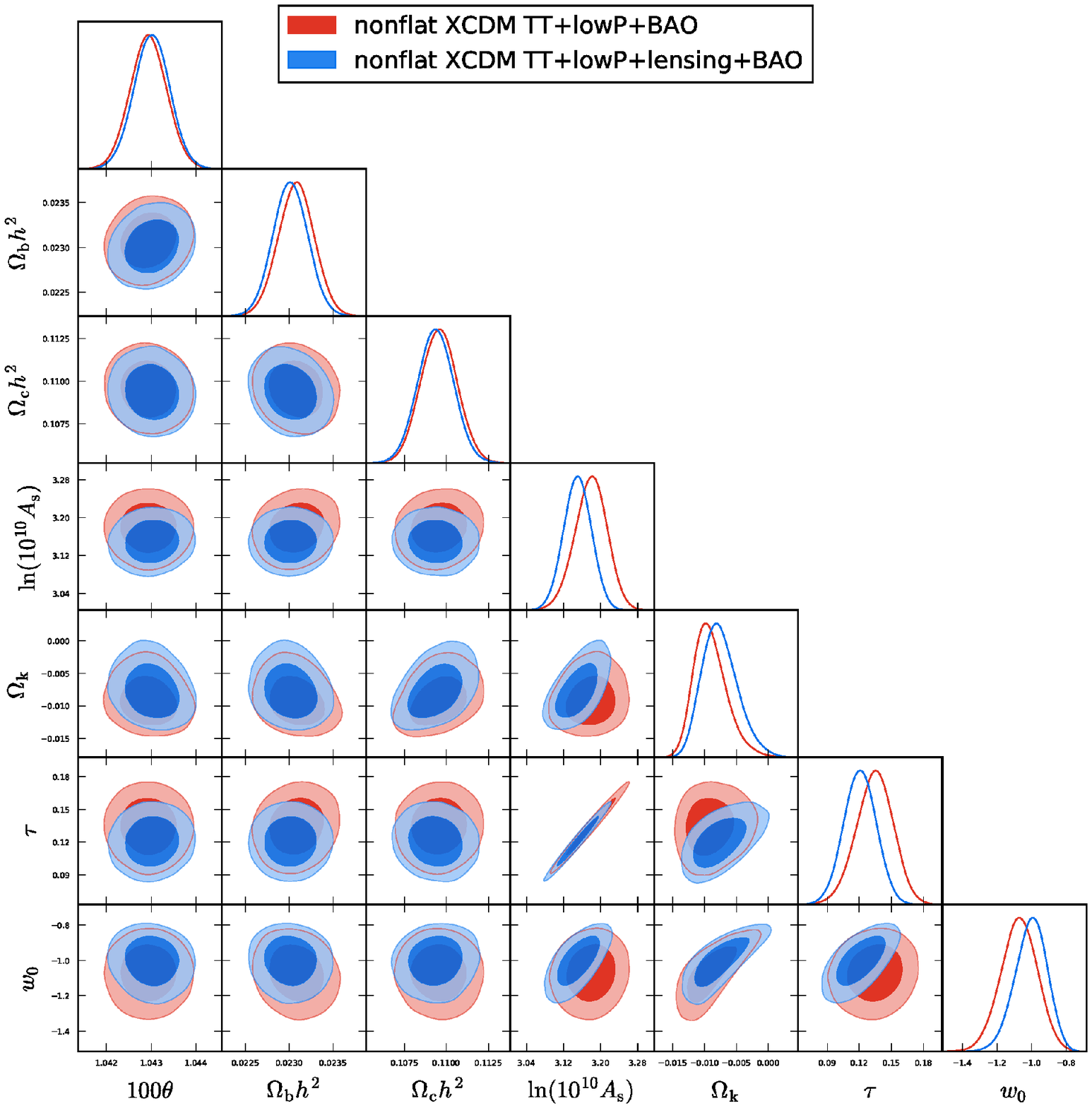}
\caption{$68.27\%$ and $95.45\%$ confidence level contours for the non-flat 
XCDM inflation model using various data sets, with the other parameters 
marginalized.\label{fig:tri}}
\end{figure}

\begin{figure}[ht]
\gridline{\fig{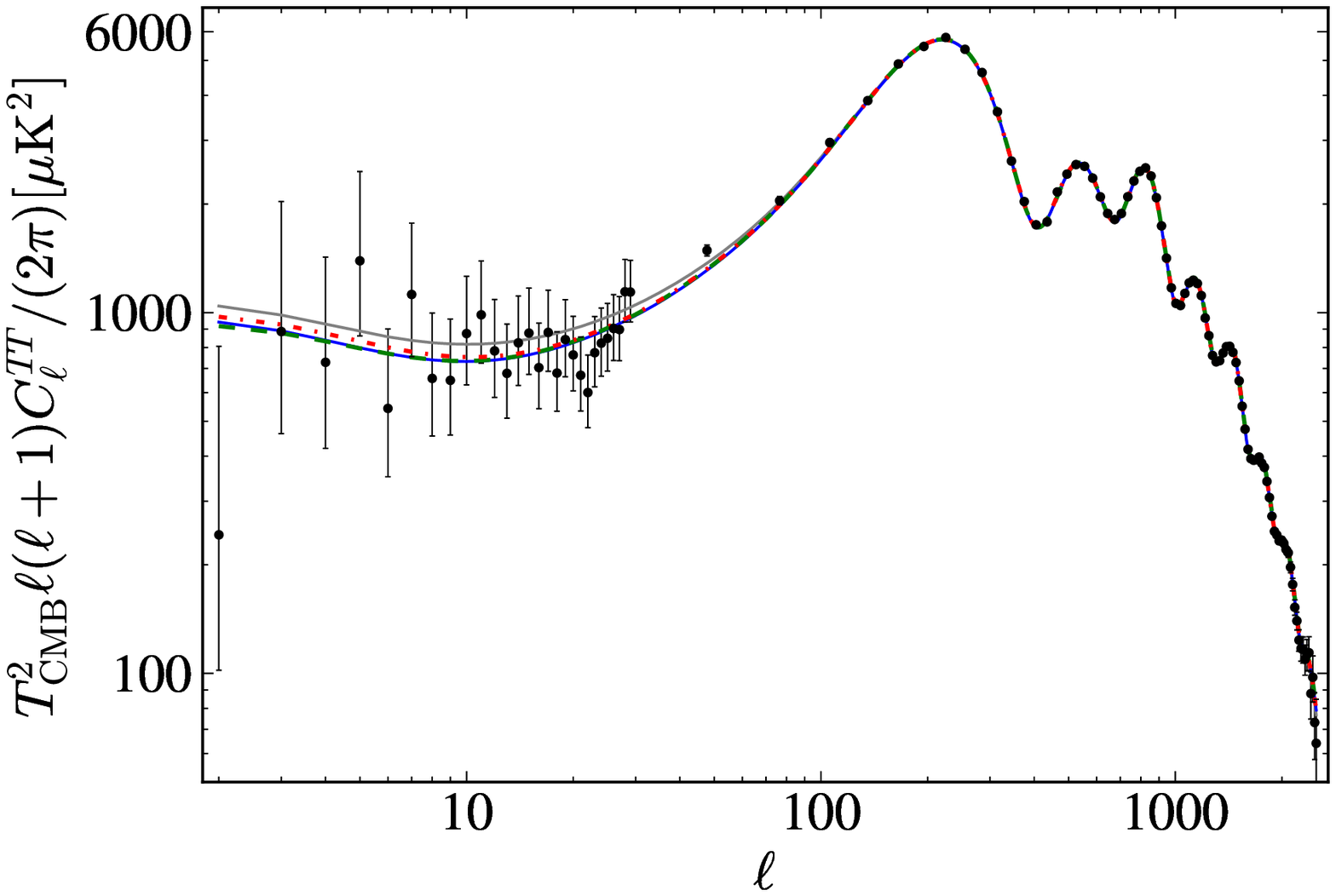}{0.45\textwidth}{(a)}
          \fig{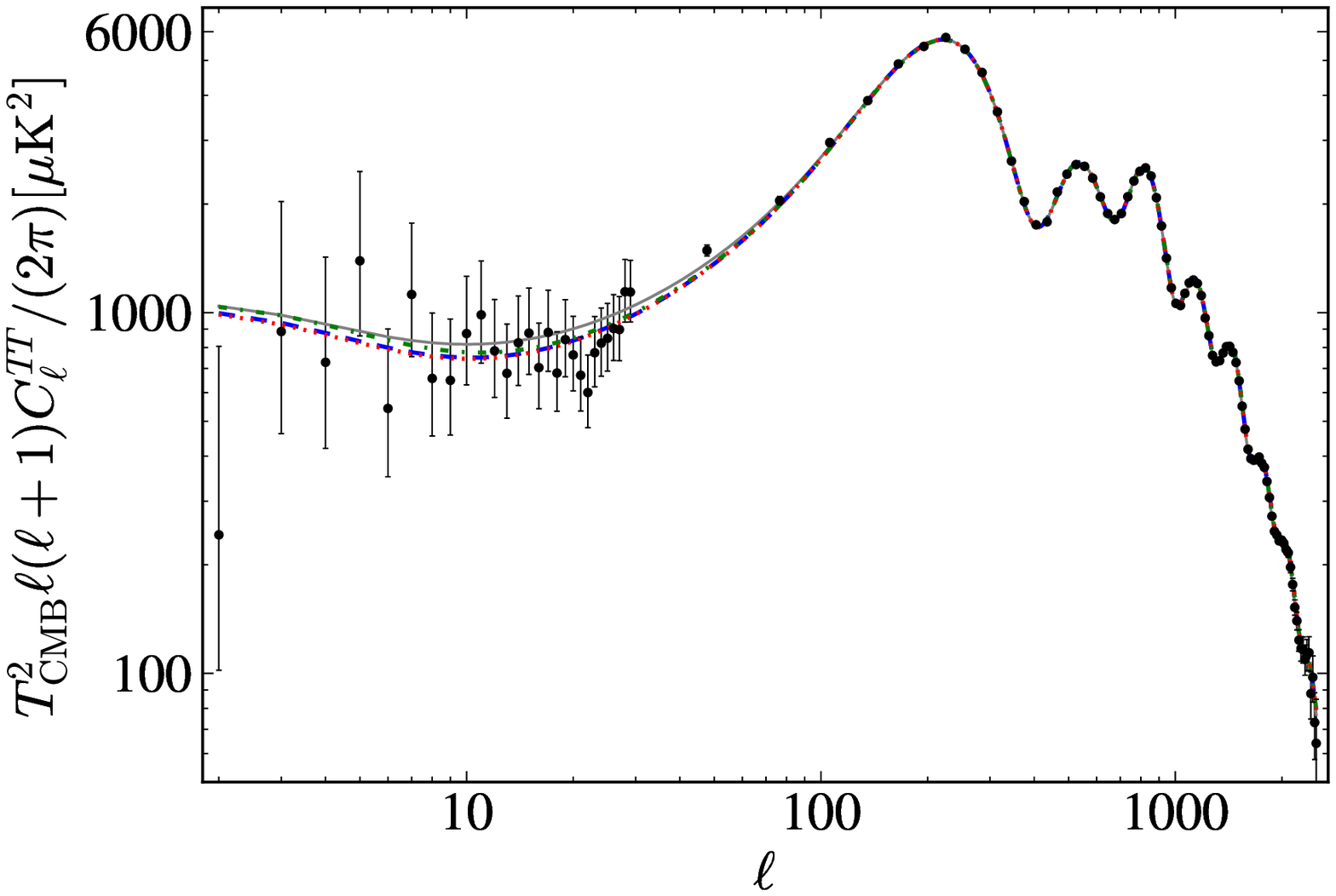}{0.45\textwidth}{(b)}
          }
\gridline{\fig{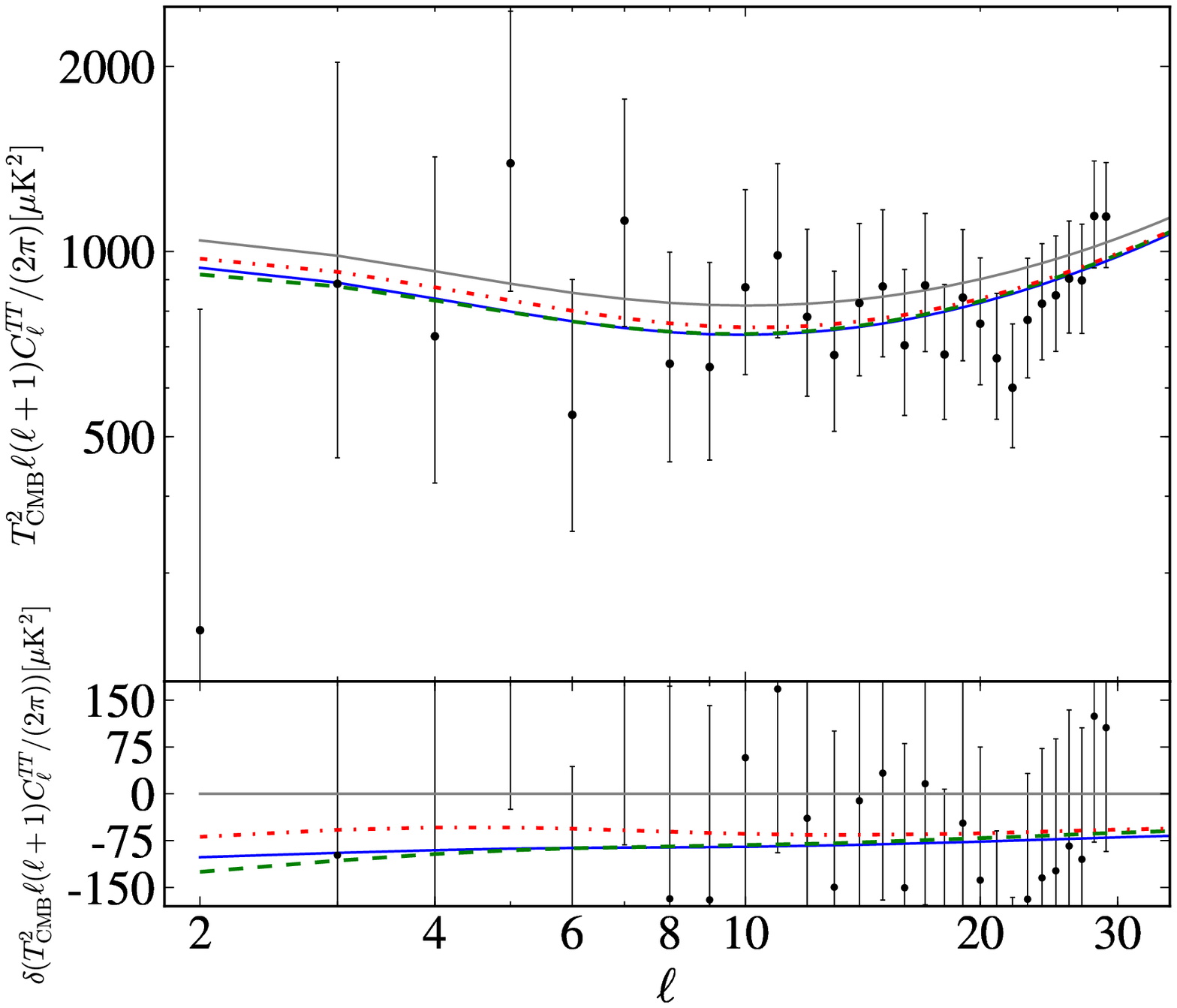}{0.42\textwidth}{(c)}
          \fig{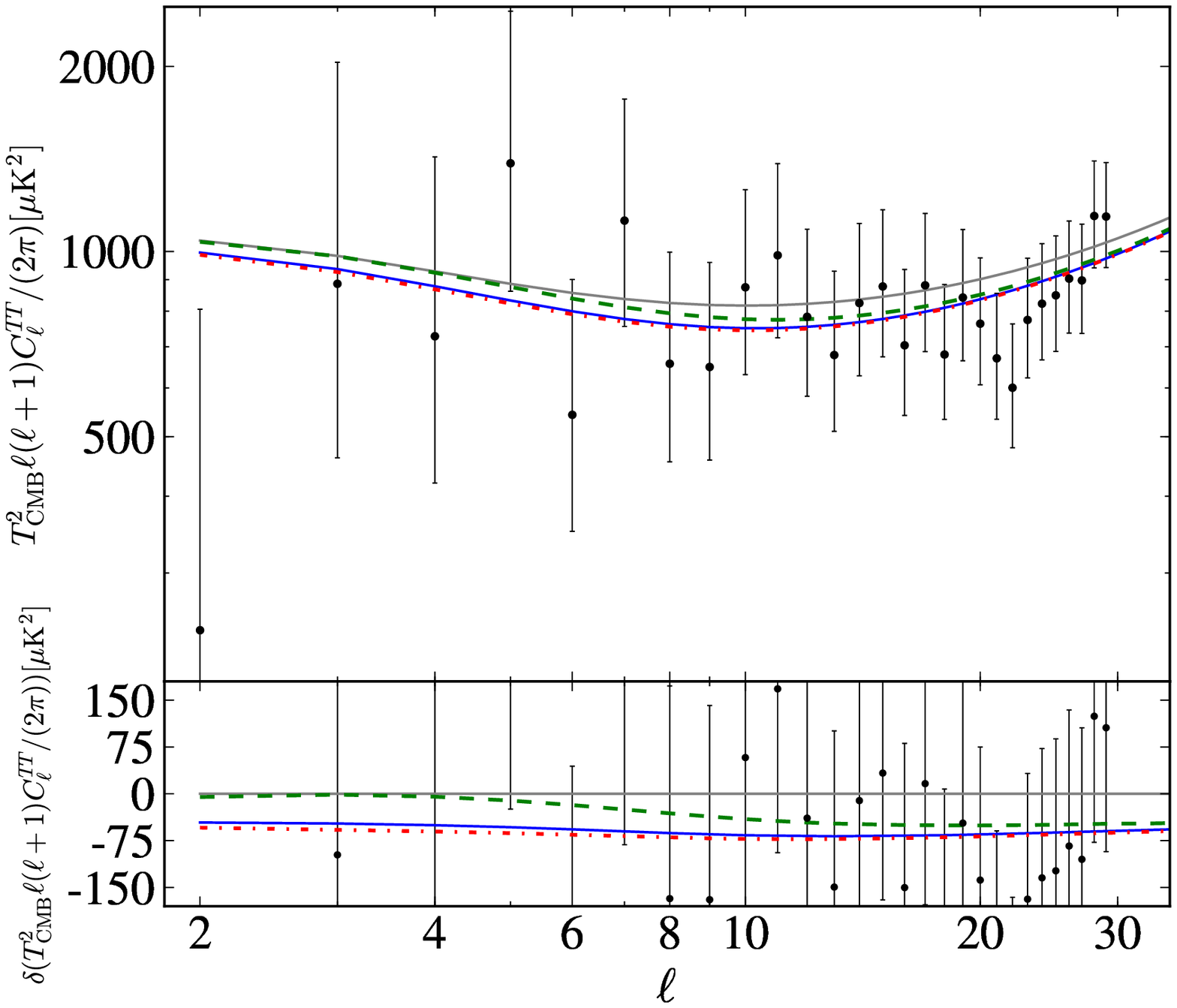}{0.42\textwidth}{(d)}
          }
\gridline{\fig{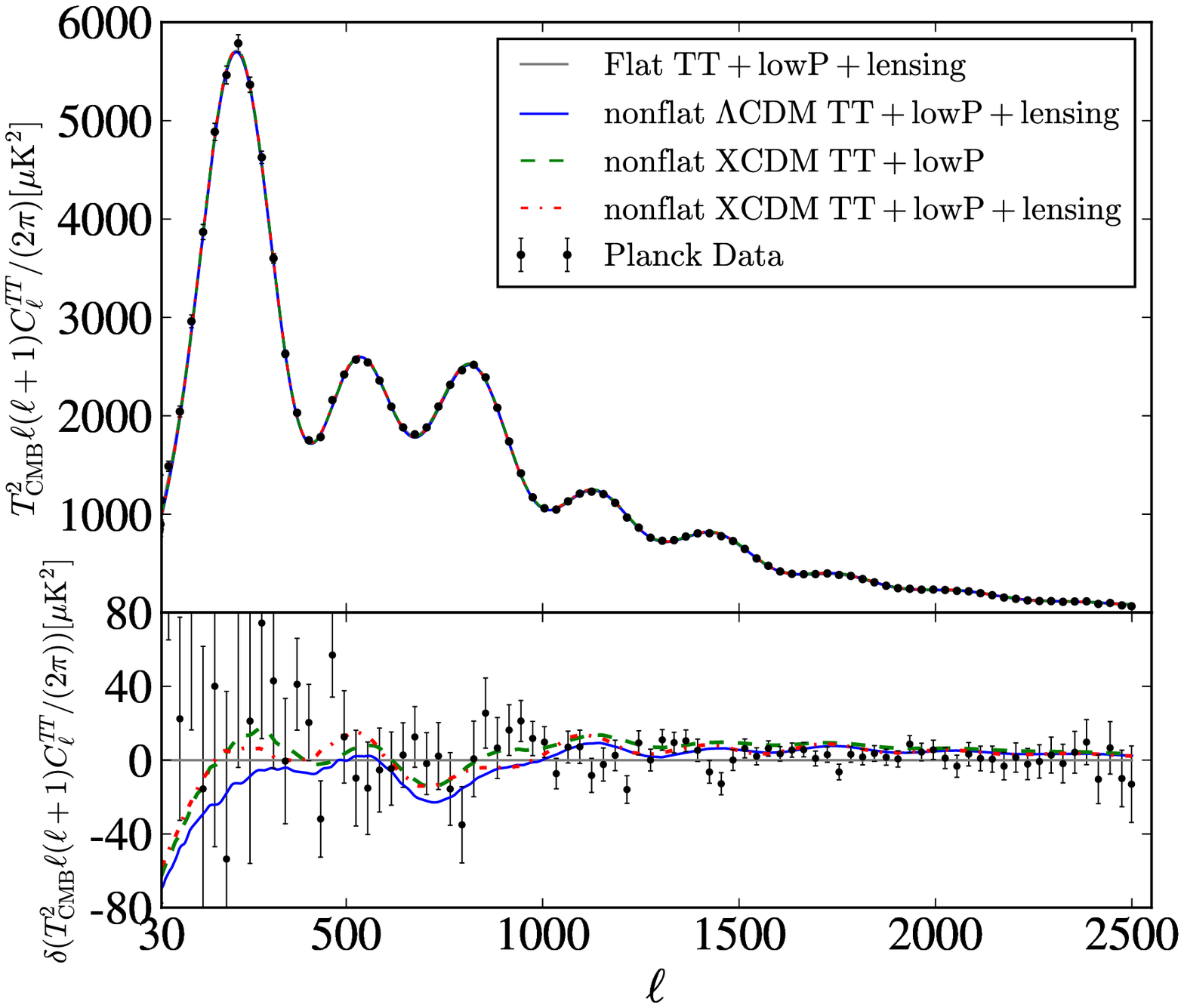}{0.43\textwidth}{(e)}
          \fig{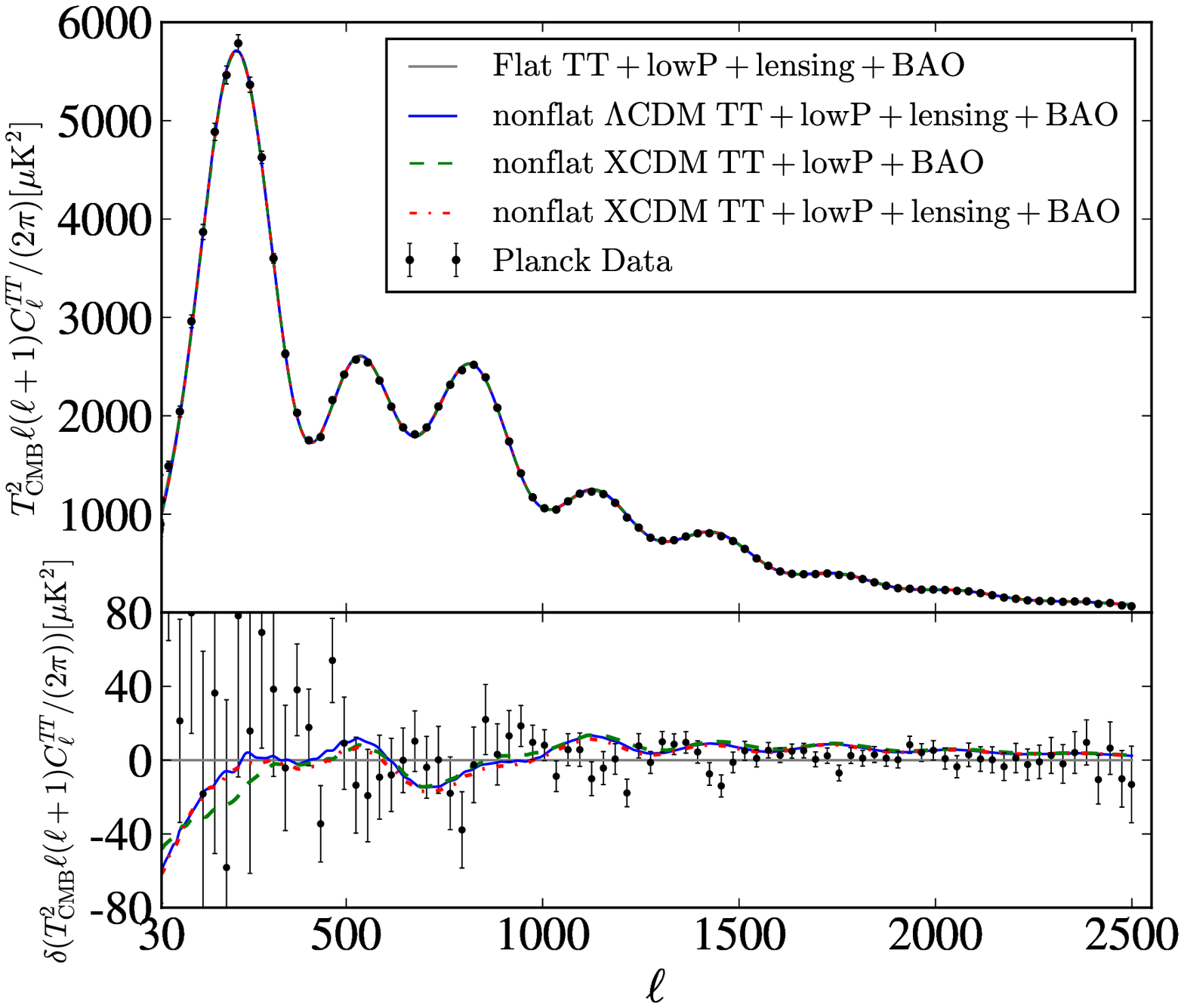}{0.43\textwidth}{(f)}
          }
\caption{Temperature $C_{\ell}$ for the best-fit non-flat XCDM, non-flat 
$\Lambda$CDM and spatially-flat tilted $\Lambda$CDM (gray solid line) models. 
Linestyle information are listed in the boxes in the two bottom panels. 
Planck 2015 measurements are shown as black points with error bars. Left 
panels (a), (c) and (e) are from CMB data alone analyses, while right 
panels (b), (d) and (f) analyses also include BAO data. Top panels show the 
all-$\ell$ region. Middle panels show the low-$\ell$ region $C_\ell$ and 
residuals. Bottom panels show the high-$\ell$ region $C_\ell$ and 
residuals.\label{fig:cls}}
\end{figure}

\begin{figure}[ht]
\plottwo{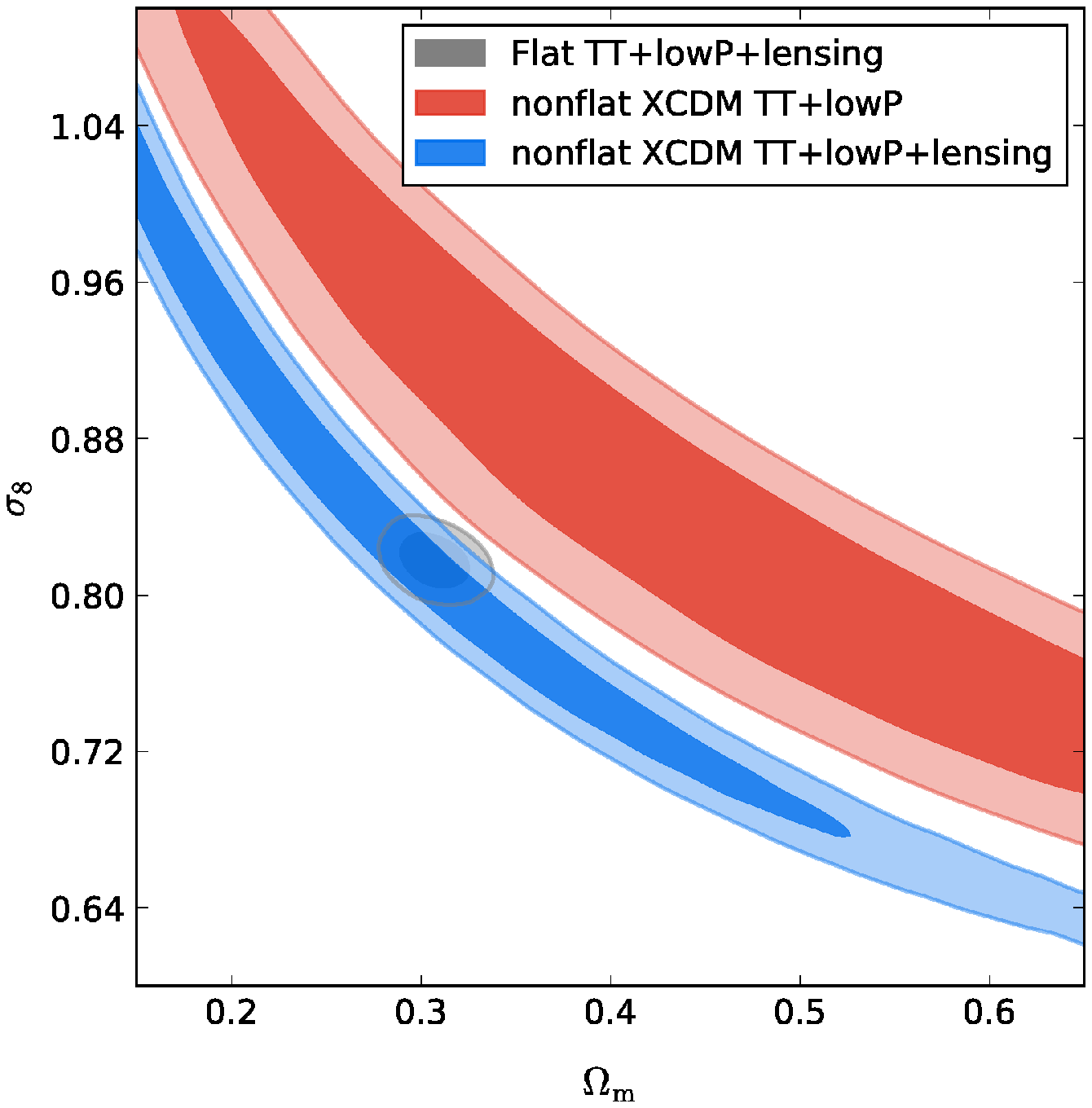}{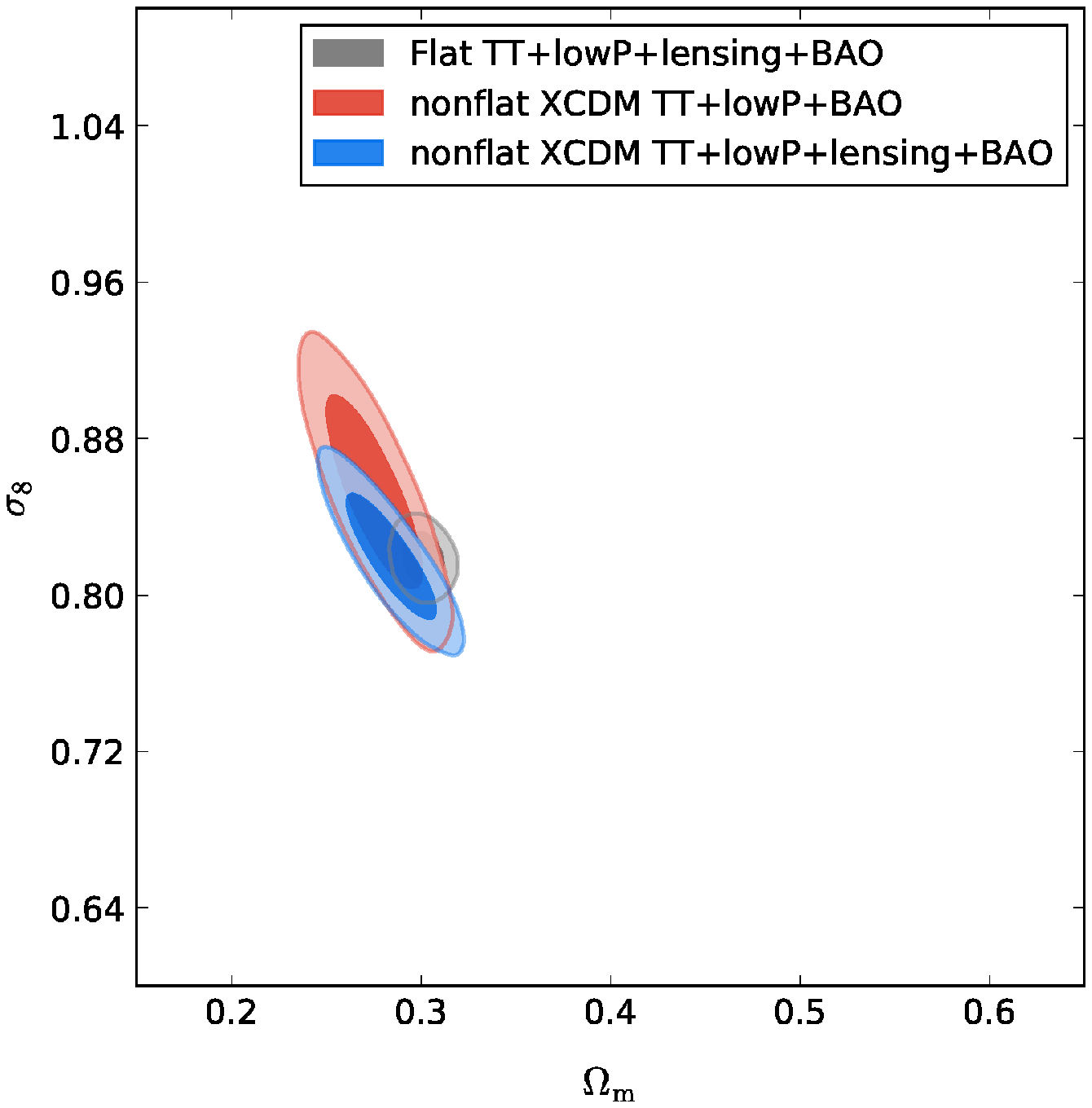}
\caption{$68.27\%$ and $95.45\%$ confidence contours in 
the $\sigma_8$--$\Omega_{\rm m}$ plane. Left panel shows Planck data alone contours while right panel contours also account for BAO data. \label{fig:sigm}}
\end{figure}

In this section we present results of our parameter estimation 
computations and our attempt to determine how well the best-fit seven parameter 
closed-XCDM inflation case does relative to the best-fit six parameter
tilted flat-$\Lambda$CDM inflation model \citep{Adeetal2016a} and the 
best-fit six parameter closed-$\Lambda$CDM inflation model 
\citep{Oobaetal2018a}.
Table \ref{tab:table1} lists central values and 
$68.27\%$ ($1\sigma$) limits on the cosmological parameters from the 
various CMB and BAO data sets we use.
Figure \ref{fig:tri} shows two-dimensional constraint contour
and one-dimensional likelihood (determined by marginalizing over all other 
parameters) plots, derived from the two CMB anisotropy data sets, both 
excluding and 
including the BAO data. Figure \ref{fig:cls} plots the CMB 
temperature anisotropy angular power spectra for the best-fit non-flat 
XCDM cases determined from the two different CMB anisotropy data sets
(as well as for the non-flat and tilted spatially-flat $\Lambda$CDM models), 
excluding and 
including the BAO data, Figure \ref{fig:sigm} shows $68.27\%$ 
and $95.45\%$ ($2\sigma$) confidence contours in the 
$\sigma_8$--$\Omega_{\rm m}$ plane, after marginalizing over the other 
parameters, for the non-flat XCDM inflation cases as well as for 
one spatially-flat tilted $\Lambda$CDM inflation model, 
without and with the BAO data.

\begin{table*}[ht]
\caption{\label{tab:table1}
68.27\% confidence limits on cosmological parameters of the non-flat XCDM model from CMB and BAO data.}
\centering
\begin{tabular}{lcccc}
\hline
\hline
\textrm{Parameter}&
\textrm{TT+lowP}&
\textrm{TT+lowP+lensing}&
\textrm{TT+lowP+BAO}&
\textrm{TT+lowP+lensing+BAO}\\
\hline
$\Omega_{\rm b}h^2$ & $0.02335\pm 0.00022$ & $0.02304\pm 0.00021$ & $0.02308\pm 0.00021$ & $0.02301\pm 0.00020$\\
$\Omega_{\rm c}h^2$ & $0.1092\pm 0.0011$ & $0.1091\pm 0.0011$ & $0.1096\pm 0.0011$ & $0.1094\pm 0.0011$\\
$100\theta$ & $1.04301\pm 0.00042$ & $1.04307\pm 0.00041$ & $1.04294\pm 0.00042$ & $1.04301\pm 0.00041$\\
$\tau$ & $0.084\pm 0.031$ & $0.100\pm 0.022$ & $0.134\pm 0.017$ & $0.121\pm 0.015$\\
${\rm ln}(10^{10}A_{\rm s})$ & $3.077\pm 0.062$ & $3.107\pm 0.044$ & $3.178\pm 0.035$ & $3.150\pm 0.030$\\
$\Omega_{\rm k}$ & $-0.086\pm 0.048$ & $-0.021\pm 0.014$ & $-0.009\pm 0.003$ & $-0.008\pm 0.003$\\
$w_0$ & $-1.55\pm 0.86$ & $-1.23\pm 0.56$ & $-1.07\pm 0.11$ & $-1.00\pm 0.10$\\
\hline
$H_0$ [km/s/Mpc] & $55.53\pm 12.76$ & $73.96\pm 20.29$ & $69.52\pm 2.16$ & $68.36\pm 1.93$\\
$\Omega_{\rm m}$ & $0.52\pm 0.24$ & $0.31\pm 0.14$ & $0.28\pm 0.02$ & $0.28\pm 0.02$\\
$\sigma_8$ & $0.829\pm 0.134$ & $0.865\pm 0.162$ & $0.852\pm 0.034$ & $0.820\pm 0.022$\\
$S_8\equiv\sigma_8\sqrt{\Omega_{\rm m}/0.3}$ & $1.020\pm 0.104$ & $0.795\pm 0.084$ & $0.815\pm 0.019$ & $0.797\pm 0.012$\\
\hline
\hline
\end{tabular}
\end{table*}

From the analysis without the BAO data, the spatial curvature 
density parameter is measured to be
\begin{equation}
\label{eq:omk}
\Omega_{\rm k} = -0.021 {+0.029 \atop -0.050} \ \ (95.45\%,\ \rm TT+lowP+lensing).
\end{equation}
The left hand panels of Fig. \ref{fig:cls} show the CMB temperature anisotropy 
$C_\ell$ of the best-fit non-flat XCDM inflation cases for the 2 
different CMB data sets. It appears that these models fit the 
low-$\ell$ $C_{\ell}$ observations better than does the spatially-flat tilted 
$\Lambda$CDM case of \citet{Adeetal2016a}, with the higher-$\ell$ $C_{\ell}$ 
data not as well fit by the non-flat models. On large scales the fractional 
energy density inhomogeneity power spectrum for the best-fit closed XCDM and 
$\Lambda$CDM inflation models are suppressed relative to that of the best-fit 
tilted flat-$\Lambda$CDM model, as shown in Fig.\ \ref{fig:Pk} for the TT + 
lowP + lensing data. The 
low-$\ell$ $C_\ell$ of Fig.\ \ref{fig:cls} depend on 
this small wavenumber part of the power spectrum, but other effects, such as 
the usual and integrated Sachs-Wolfe effects, also play an important role 
in determining the $C_\ell$ shape. The left panel of Fig. 
\ref{fig:sigm} shows $\sigma_8$--$\Omega_{\rm m}$ constraint contours
for the 2 non-flat XCDM cases (as well as for one spatially-flat 
tilted $\Lambda$CDM model). With CMB lensing included, we find that our 
non-flat seven parameter XCDM inflation model reduces the tension 
between the CMB observations and the weak lensing data, compare Fig.\ 
\ref{fig:sigm} here 
to Fig. 18 of \citet{Adeetal2016a}.\footnote{Note that, as discussed in Sec.\ 5 of \citet{Adeetal2016a}, because of the incorrect amount of lensing in the TT power spectrum, this tension is also alleviated when the Planck lensing reconstruction data is used \citep{Renzietal2018}. For another option see \citet{DiValentinoetal2018}.}  

From the analysis also including the BAO data, the spatial 
curvature density parameter is measured to be
\begin{equation}
\label{eq:omk2}
\Omega_{\rm k} = -0.008 \pm 0.006\ \ (95.45\%,\ \rm TT+lowP+lensing+BAO).
\end{equation}
In contrast to the Planck 2015 results \citep{Adeetal2016a}, our 
physically-consistent non-flat XCDM inflation model is not forced to be flat 
even with the BAO data included in the analysis. This case is about 3$\sigma$ 
away from flat. The right panels of Fig. \ref{fig:cls} show temperature 
$C_{\ell}$ plots for the 
best-fit non-flat XCDM models analyzed using the 2 different CMB
data sets and including the BAO data. Including the BAO data 
does somewhat degrade the fit in the low-$\ell$ region compared with results 
from the analyses without the BAO data. We find that including the BAO data 
also worsens the 
$\sigma_8$--$\Omega_{\rm m}$ plane tension between the CMB and weak lensing
constraints, see the right hand panel of Fig. \ref{fig:sigm}.

It is interesting that the $\Omega_{\rm m}$ and $H_0$ constraints listed in 
Table \ref{tab:table1} are very consistent with estimates for these 
parameters from most other data. See \citet{ChenRatra2003} for the density 
parameter. The most recent median statistics analysis of $H_0$ 
measurements gives $H_0 = 68 \pm 2.8$ km s${}^{-1}$ Mpc${}^{-1}$ 
\citep{ChenRatra2011}, consistent with earlier estimates \citep{Gottetal2001, Chenetal2003}. Many recent $H_0$ determinations from BAO, Type Ia supernovae, 
Hubble parameter, and other measurements are consistent with these results
\citep{Calabreseetal2012, Hinshawetal2013, Sieversetal2013, Aubourgetal2015, LHuillierShafieloo2017, Lukovicetal2016, Chenetal2017, Wangetal2017, LinIshak2017}. It is however well known 
that local measurements of the expansion 
rate find a higher $H_0$ value. \citet{Freedmanetal2012} give 
$H_0 = 74.3 \pm 2.1$ km s${}^{-1}$ Mpc${}^{-1}$ while \citet{Riessetal2016} 
report $H_0 = 73.24 \pm 1.74$ km s${}^{-1}$ Mpc${}^{-1}$, somewhat higher than the 
$H_0 = 68.36 \pm 1.93$ km s${}^{-1}$ Mpc${}^{-1}$ of the last column of Table 
\ref{tab:table1}. 

In addition, we note that many analyses based on a number of 
different observations also do not rule out non-flat dark energy models
\citep{Farooqetal2015, Saponeetal2014, Lietal2014, Caietal2016, Chenetal2016, YuWang2016, LHuillierShafieloo2017, Farooqetal2017, Lietal2016, WeiWu2017, Ranaetal2017, Yuetal2018, Mitraetal2018, Ryanetal2018}.

\begin{table*}[ht]
\caption{\label{tab:table2}
Minimum $\chi^2$ values for the best-fit closed-XCDM (and tilted flat-$\Lambda$CDM)
inflation model.}
\centering
\begin{tabular}{lcc}
\hline
\hline
\textrm{Data sets}& \textrm{$\chi^2$}& \textrm{d.o.f.}\\
\hline
TT+lowP & 11277 (11262) & 188 (189)\\
TT+lowP+lensing & 11292 (11272) & 196 (197)\\
TT+lowP+BAO & 11289 (11266) & 192 (193)\\
TT+lowP+lensing+BAO & 11298 (11277) & 200 (201)\\
\hline
\hline
\end{tabular}
\end{table*}

\begin{table*}[ht]
\caption{\label{tab:table3}
$\chi^2$ values for the best-fit closed-XCDM (and tilted flat-$\Lambda$CDM)
inflation model.}
\centering
\begin{tabular}{lcccc}
\hline
\hline
\textrm{Data}&
\textrm{TT+lowP}&
\textrm{TT+lowP+lensing}&
\textrm{TT+lowP+BAO}&
\textrm{TT+lowP+lensing+BAO}\\
\hline
CMB low-$\ell$ & 130.15 (129.83) & 131.39 (126.06) & 140.07 (130.06) & 129.78 (126.00)\\
CMB high-$\ell$ & 138.86 (77.13) & 121.31 (82.46) & 143.45 (73.64)  & 134.92 (80.82)\\
CMB all-$\ell$ & 269.01 (206.97) & 262.99 (218.43) & 286.44 (207.73) & 279.18 (220.31)\\
CMB lensing & --- & 9.81 (9.90) & --- & 9.77 (9.92)\\
BAO & --- & --- & 2.92 (4.03) & 4.70 (3.58)\\
\hline
d.o.f.\ & 188 (189) & 196 (197) & 192 (193) & 200 (201) \\
\hline
\hline
\end{tabular}
\end{table*}

It is vital to understand how well the best-fit closed-XCDM 
inflation case does relative to the best-fit tilted flat-$\Lambda$CDM model in
fitting the data. As for the closed-$\Lambda$CDM model \citep{Oobaetal2018a}, 
we are unable to resolve this in a quantitative manner, although qualitatively,
overall, the best-fit closed-XCDM model does not do as well as the best-fit 
tilted flat-$\Lambda$CDM model. It also appears to not do as well as the
best-fit closed-$\Lambda$CDM inflation model \citep{Oobaetal2018a}, which has 
one fewer parameter.

Table \ref{tab:table2} lists the minimum $\chi^2 = -2 {\rm ln} 
({\rm L_{\rm max}})$ determined from the maximum value of the likelihood, for 
the 4 data sets we consider, for both the closed-XCDM and tilted
flat-$\Lambda$CDM inflation models, as well as the number of (binned data) 
degrees of freedom (d.o.f.). The d.o.f.\ are determined from combinations of
112 low-$\ell$ TT + lowP, 83 high-$\ell$ TT, 8 lensing CMB (binned) 
measurements, 4 BAO observations, and 7 (or 6 for the tilted flat-$\Lambda$CDM)
model parameters. It is likely that the large $\chi^2$ values are the result 
of the many nuisance parameters that have been marginalized over, since 
the tilted flat-$\Lambda$CDM model is said to be a good fit to the data.
From this table we see that $\Delta \chi^2 = 20 (21)$ for the 
closed-XCDM inflation case (196 (200) d.o.f.),
relative to the tilted flat-$\Lambda$CDM (197 (201) d.o.f.), 
for the TT + low P + lensing (+ BAO) data. This might make the 
closed-XCDM case much less probable, however, one can see from the residual 
panels of Fig. \ref{fig:cls} (e) \& (f) that this $\Delta \chi^2$ appears 
to be a consequences of  many small deviations, 
and not of a few significant outliers. 

Though there are correlations in the data, it is useful to also consider 
a standard goodness of fit $\chi^2$ that makes use of only the diagonal 
elements of the correlation matrix. These $\chi^2$'s are listed in 
Table \ref{tab:table3} for the 4 data sets we study and for both the 
closed-XCDM and tilted flat-$\Lambda$CDM inflation models. From Table
\ref{tab:table3} for the TT + low P + lensing data, the $\chi^2$ 
per d.o.f.\ is 273/196 (227/197) for the closed-XCDM (tilted 
flat-$\Lambda$CDM) inflation model, and when 
BAO data is included these become 294/200 (234/201). Again, 
while the closed-XCDM case is less favored than the tilted 
flat-$\Lambda$CDM case, it is not clear how to assess the quantitative
significance of this. In addition to the discussion at the end of the 
previous paragraph, here we have also ignored all off-diagonal 
information in the correlation matrix, so it is meaningless to compute 
standard probabilities from such $\chi^2$'s. In summary, while the best-fit 
closed-XCDM inflation case appears less favored, it might be useful 
to perform a more thorough analysis of the model.

We additionally compute the Bayesian evidence for each of the models we 
consider, by using the public numerical code MCEvidence \citep{Heavensetal2017}.
We then compute the natural logarithm of the ratio of the Bayesian evidence 
of our model to that of the spatially-flat tilted $\Lambda$CDM model, which we 
write as $\ln(B)$. For the TT + low P (+ lensing) case, we get 
$\ln(B) = +4.08\ (-3.79)$, and for the TT + low P + BAO (+ lensing) case, we 
find $\ln(B) = -14.98\ (-18.51)$. According to \citet{KassRaftery1995}, 
$|\ln B| > 3$ is strong evidence and $|\ln B| > 5$ is very strong.
So our model is favored when we use the TT + low P data, while it is 
less favored than tilted flat-$\Lambda$CDM when we consider the other data sets.

\section{Conclusion}

We determine Planck 2015 CMB data constraints on the physically consistent 
seven parameter non-flat XCDM model with inflation-generated 
energy density inhomogeneity power spectrum. This is a first 
attempt to examine the effect of the interplay between dark energy dynamics
and spatial curvatures on constraints derived from CMB anisotropy measurements.
We again draw attention to the fact that the X-fluid dark energy part of the 
XCDM model is only a parameterization, and not a physical model, and hope 
it does not lead to misleading conclusions.

Unlike the case for 
the seven parameter non-flat tilted $\Lambda$CDM model with the 
physically-inconsistent power 
spectrum used in \citet{Adeetal2016a}, we discover that CMB anisotropy data 
do not force spatial curvature to vanish in our non-flat XCDM inflation model
with a physically-consistent power spectrum. 
This is consistent with what we found in the six parameter non-flat 
$\Lambda$CDM inflation model \citep{Oobaetal2018a}, where spatial 
curvature contributes about 2 \% to the present energy budget of the closed 
model that best fits the Planck TT + lowP + lensing data. These closed 
inflation models are more consistent with the low-$\ell$ $C_{\ell}$ temperature 
observations and the weak lensing 
$\sigma_8$ measurements than is the best fit spatially-flat tilted $\Lambda$CDM,
but they do worse at fitting the higher-$\ell$ ${C_{\ell}}$ 
data.\footnote{Deuterium abundance measurements mildly favor the flat case over 
the open one \citep{Pentonetal2018} while a compilation of all recent BAO, Type Ia supernova,
and Hubble parameter data mildly favor closed spatial hypersurfaces \citep{ParkRatra2018d}.}

It might be useful to reexamine the issue of possible small differences in 
cosmological parameter constraints derived from higher-$\ell$ and from 
lower-$\ell$ Planck 2015 CMB 
anisotropy data \citep{Addisonetal2016, Aghanimetal2016}, by using the 
non-flat XCDM model. In addition we note that in the tilted flat $\Lambda$CDM model
there seem to be inconsistencies between the higher-$\ell$ Planck and the South Pole Telescope CMB data \citep{Ayloretal2017}.
Also of great interest would be a method for 
quantitatively assessing how well the 
best-fit tilted spatially-flat $\Lambda$CDM model and the best-fit non-flat 
XCDM case fit the Planck 2015 CMB anisotropy (and other cosmological) 
measurements.

Unlike the analysis for the seven parameter non-flat tilted $\Lambda$CDM model 
in \citet{Adeetal2016a}, also including the BAO data in the mix does not force 
our seven parameter non-flat XCDM case to be flat;
in fact, $\Omega_{\rm k} =  -0.008 \pm 0.006$ at 2$\sigma$ and is about 
3$\sigma$ away from flat. In this case the improved agreement with the 
low-$\ell$ $C_{\ell}$ temperature data and the weak lensing $\sigma_8$ 
constraints are not as good compared with the results from the analyses 
using only the CMB observations. We note that the BAO and CMB data are 
from very different redshifts 
and it is possible that a better model for the intervening epoch or an improved 
understanding of one or both sets of observations might alter this finding.

Perhaps adding a small spatial curvature contribution, of order a percent, can 
improve the spatially-flat standard $\Lambda$CDM model.
However, it appears that the six parameter closed-$\Lambda$CDM inflation 
model might do a better job at this \citep{Oobaetal2018a, ParkRatra2018a, ParkRatra2018b} than does the 
seven parameter closed-XCDM inflation model we studied here. A more thorough 
analysis of the non-flat $\Lambda$CDM and XCDM inflation models is needed to 
establish if either is viable and can help resolve some of 
the low-$\ell$ $C_{\ell}$ issues as well as possibly the $\sigma_8$ power 
issues. Perhaps non-zero spatial curvature might be more important for 
this purpose than is dark energy dynamics. 

\section*{Acknowledgments}

WE acknowledge valuable discussions with C.-G.\ Park. This work is supported 
by Grants-in-Aid for Scientific Research 
from JSPS (Nos.\ 16J05446 (J.O.) and 15H05890 (N.S.)). B.R.\ is supported 
in part by DOE grant DE-SC0019038.

\newpage


\end{document}